\documentclass{article}

\usepackage[nonatbib,final]{neurips_2019}

\usepackage[
backend=biber,
style=numeric,
citestyle=numeric-comp
]{biblatex}

\usepackage[utf8]{inputenc} 
\usepackage[T1]{fontenc}    
\usepackage{hyperref}       
\usepackage{url}            
\usepackage{booktabs}       
\usepackage{amsfonts}       
\usepackage{nicefrac}       
\usepackage{microtype}      

\usepackage{makecell}

\usepackage{xcolor}

\usepackage{graphicx}
\usepackage{enumitem}

\title{Women, politics and Twitter: \\ Using machine learning to change the discourse}

\author{%
  Lana Cuthberston\thanks{Corresponding author} \\
  ParityYEG\\
  \texttt{lanacuthbertson@gmail.com} \\
  \And
  Alex Kearney \\
  University of Alberta \\
  \texttt{kearney@ualberta.ca} \\
   \And
   Riley Dawson \\
   University of Alberta \\
   \texttt{rileydawson@ualberta.ca} \\
   \And
   Ashia Zawaduk \\
   Dev Edmonton Society \\
   \texttt{mail@ashia.ca} \\
   \And
   Eve Cuthbertson \\
   University of Toronto \\
   \texttt{evecuthbertson@gmail.com } \\
   \And
   Ann Gordon-Tighe \\
   University of Alberta \\
   \texttt{ann.gordontighe@gmail.com} \\
   \AND
   Kory W Mathewson \\
   University of Alberta \\
   \texttt{korymath@google.com} \\
}

\begin{document}

\maketitle

\begin{abstract}
Including diverse voices in political decision-making strengthens our democratic institutions. Within the Canadian political system, there is gender inequality across all levels of elected government. Online abuse, such as hateful tweets, leveled at women engaged in politics contributes to this inequity, particularly tweets focusing on their gender. In this paper, we present \textit{ParityBOT}: a Twitter bot which counters abusive tweets aimed at women in politics by sending supportive tweets about influential female leaders and facts about women in public life. ParityBOT is the first artificial intelligence-based intervention aimed at affecting online discourse for women in politics for the better. The goal of this project is to: $1$) raise awareness of issues relating to gender inequity in politics, and $2$) positively influence public discourse in politics. The main contribution of this paper is a scalable model to classify and respond to hateful tweets with quantitative and qualitative assessments. The ParityBOT abusive classification system was validated on public online harassment datasets. We conclude with analysis of the impact of ParityBOT, drawing from data gathered during interventions in both the $2019$ Alberta provincial and $2019$ Canadian federal elections.

\end{abstract}

\section{Introduction}

Our political systems are unequal, and we suffer for it. Diversity in representation around decision-making tables is important for the health of our democratic institutions \cite{phillips1998democracy}. One example of this inequity of representation is the gender disparity in politics: there are fewer women in politics than men, largely because women do not run for office at the same rate as men. This is because women face systemic barriers in political systems across the world \cite{ipu}. One of these barriers is online harassment \cite{trimble2018ms,rheault2019politicians}. Twitter is an important social media platform for politicians to share their visions and engage with their constituents. Women are disproportionately harassed on this platform because of their gender \cite{toxictwitter}. 

To raise awareness of online abuse and shift the discourse surrounding women in politics, we designed, built, and deployed \textit{ParityBOT}: a Twitter bot that classifies hateful tweets directed at women in politics and then posts ``positivitweets''. This paper focuses on how ParityBOT improves discourse in politics.

Previous work that addressed online harassment focused on collecting tweets directed at women engaged in politics and journalism and determining if they were problematic or abusive \cite{delisle2019large,rheault2019politicians,greenwood2019online}. Inspired by these projects, we go one step further and develop a tool that directly engages in the discourse on Twitter in political communities. Our hypothesis is that by seeing ``positivitweets'' from ParityBOT in their Twitter feeds, knowing that each tweet is an anonymous response to a hateful tweet, women in politics will feel encouraged and included in digital political communities\cite{Africaworkshop}. This will reduce the barrier to fair engagement on Twitter for women in politics. It will also help achieve gender balance in Canadian politics and improve gender equality in our society.

\section{Methods}

\subsection{Technical Details for ParityBot}
\label{sec:tech}

In this section, we outline the technical details of ParityBot. The system consists of: $1$) a Twitter listener that collects and classifies tweets directed at a known list of women candidates, and $2$) a responder that sends out positivitweets when hateful tweets are detected. 


We collect tweets from Twitter's real-time streaming API. The stream listener uses the open-source Python library Tweepy \cite{tweepy}. The listener analyses tweets in real-time by firing an asynchronous tweet analysis and storage function for each English tweet mentioning one or more candidate usernames of interest. We limit the streaming to English as our text analysis models are trained on English language corpora. We do not track or store retweets to avoid biasing the analysis by counting the same content multiple times. Twitter data is collected and used in accordance with the acceptable terms of use \cite{twitterdev}. 

The tweet analysis and storage function acts as follows: $1$) parsing the tweet information to clean and extract the tweet text, $2$) scoring the tweet using multiple text analysis models, and $3$) storing the data in a database table. We clean tweet text with a variety of rules to ensure that the tweets are cleaned consistent with the expectations of the analysis models (see Appdx \ref{appdx:clean}).

The text analysis models classify a tweet by using, as features, the outputs from Perspective API from Jigsaw \cite{perspective}, HateSonar \cite{hatesonar}, and VADER sentiment models \cite{hutto2014vader}. Perspective API uses machine learning models to score the perceived impact a tweet might have \cite{perspective}. The outputs from these models (i.e. $17$ from Perspective, $3$ from HateSonar, and $4$ from VADER) are combined into a single feature vector for each tweet (see Appdx \ref{appdx:featz}). No user features are included in the tweet analysis models. While these features may improve classification accuracy they can also lead to potential bias \cite{waseem2016hateful}.

We measure the relative correlation of each feature with the hateful or not hateful labels. We found that Perspective API's \textsc{TOXICITY} probability was the most consistently predictive feature for classifying hateful tweets. Fig. \ref{fig:kde} shows the relative frequencies of hateful and non-hateful tweets over \textsc{TOXICITY} scores. During both elections, we opted to use a single Perspective API feature to trigger sending positivitweets. Using the single \textsc{TOXICITY} feature is almost as predictive as using all features and a more complex model \ref{appdx:ablate}. It was also simpler to implement and process tweets at scale. The \textsc{TOXICITY} feature is the only output from the Perspective API with transparent evaluation details summarized in a Model Card \cite{mitchell2019model,toxiccard}.

\subsection{Collecting Twitter handles, predicting candidate gender, curating ``positivitweets''}

Deploying ParityBOT during the Alberta $2019$ election required volunteers to use online resources to create a database of all the candidates running in the Alberta provincial election. Volunteers recorded each candidate's self-identifying gender and Twitter handle in this database. For the $2019$ federal Canadian election, we scraped a Wikipedia page that lists candidates \cite{wiki}.
We used the Python library \textit{gender-guesser} \cite{gender} to predict the gender of each candidate based on their first names. As much as possible, we manually validated these predictions with corroborating evidence found in candidates' biographies on their party's websites and in their online presence.  

ParityBOT sent positivitweets composed by volunteers. These tweets expressed encouragement, stated facts about women in politics, and aimed to inspire and uplift the community. Volunteers submitted many of these positivitweets through an online form.\footnote{The full list of positivitweets can be found at \url{https://paritybot.com}} Volunteers were not screened and anyone could access the positivitweet submission form. However, we mitigate the impact of trolls submitting hateful content, submitter bias, and ill-equipped submitters by reviewing, copy editing, and fact checking each tweet. Asking for community contribution in this way served to maximize limited copywriting resources and engage the community in the project.  

\subsection{Qualitative Assessment}

We evaluated the social impact of our system by interviewing individuals involved in government ($n=5$). We designed a discussion guide based on user experience research interview standards to speak with politicians in relevant jurisdictions \cite{krug2009rocket}. Participants had varying levels of prior awareness of the ParityBOT project. Our participants included $3$ women candidates, each from a different major political party in the $2019$ Alberta provincial election, and $2$ men candidates at different levels of government representing Alberta areas. The full discussion guide for qualitative assessment is included in Appdx \ref{appdx:disc}. All participants provided informed consent to their anonymous feedback being included in this paper.

\section{Results and Outcomes}
We deployed ParityBOT during two elections: 1) the $2019$ Alberta provincial election, and 2) the $2019$ Canadian federal election. For each tweet we collected, we calculated the probability that the tweet was hateful or abusive. If the probability was higher than our response decision threshold, a positivitweet was posted. Comprehensive quantitative results are listed in Appendix \ref{appdx:results}. 

\begin{figure}
    \centering
    \includegraphics[scale=0.4]{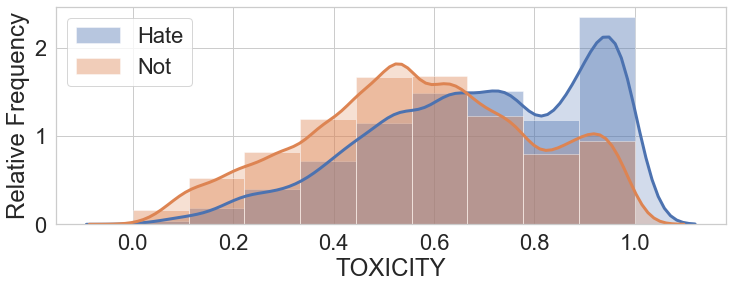}
    \caption{Visualizing the training data distribution. Relative frequency of hateful versus not hateful tweets for varying levels of the Perspective API \cite{perspective} \textsc{TOXICITY} score. Normalized histograms are plotted underneath kernel density estimation (KDE) plots.}
    \label{fig:kde}
\end{figure}

During the Alberta election, we initially set the decision threshold to a \textsc{TOXICITY} score above $0.5$ to capture the majority of hateful tweets, but we were sending too many tweets given the number of positivitweets we had in our library and the Twitter API daily limit \cite{twitterdev}. Thus, after the first $24$ hours that ParityBOT was live, we increased the decision threshold to $0.8$, representing a significant inflection point for hatefulness in the training data (Fig. \ref{fig:kde}). We further increased the decision threshold to $0.9$ for the Canadian federal election given the increase in the number and rate of tweets processed. For the Alberta provincial election, the model classified $1468$ tweets of the total $12726$ as hateful, and posted only $973$ positivitweets. This means that we did not send out a positivitweet for every classified hateful tweet, and reflects our decision rate-limit of ParityBOT. Similar results were found for the $2019$ Canadian election.

\subsection{Values and Limitations}

We wrote guidelines and values for this to guide the ongoing development of the ParityBOT project.\footnote{\url{https://paritybot.com}} These values help us make decision and maintain focus on the goal of this project.

While there is potential to misclassify tweets, the repercussions of doing so are limited. With ParityBOT, false negatives, hateful tweets classified as non-hateful, are not necessarily bad, since the bot is tweeting a positive message. False positives, non-hateful tweets classified as hateful, may result in tweeting too frequently, but this is mitigated by our choice of decision threshold. 

In developing ParityBOT, we discussed the risks of using bots on social media and in politics. First, we included the word ``bot'' in the project title and Twitter handle to be clear that the Twitter account was tweeting automatically. We avoided automating any direct ``at (@) mention'' of Twitter users, only identifying individuals' Twitter handles manually when they had requested credit for their submitted positivitweet. We also acknowledge that we are limited in achieving certainty in assigning a gender to each candidate. 

\subsection{User experience research results}

In our qualitative research, we discovered that ParityBOT played a role in changing the discourse. One participant said, ``it did send a message in this election that there were people watching'' (P2). We consistently heard that negative online comments are a fact of public life, even to the point where it's a signal of growing influence. ``When you're being effective, a good advocate, making good points, people are connecting with what you're saying. The downside is, it comes with a whole lot more negativity [\dots] I can always tell when a tweet has been effective because I notice I'm followed by trolls'' (P1). 

We heard politicians say that the way they have coped with online abuse is to ignore it. One participant explained, ``I've tried to not read it because it's not fun to read horrible things about yourself'' (P4). Others dismiss the idea that social media is a useful space for constructive discourse: ``Because of the diminishing trust in social media, I'm stopping going there for more of my intelligent discourse. I prefer to participate in group chats with people I know and trust and listen to podcasts'' (P3). 


\section{Future Work and Conclusions}

We would like to run ParityBOT in more jurisdictions to expand the potential impact and feedback possibilities. In future iterations, the system might better match positive tweets to the specific type of negative tweet the bot is responding to. Qualitative analysis helps to support the interventions we explore in this paper. To that end, we plan to survey more women candidates to better understand how a tool like this impacts them. Additionally, we look forward to talking to more women interested in politics to better understand whether a tool like this would impact their decision to run for office. 
We would like to expand our hateful tweet classification validation study to include larger, more recent abusive tweet datasets \cite{basile2019semeval,founta2018large}. We are also exploring plans to extend ParityBOT to invite dialogue: for example, asking people to actively engage with ParityBOT and analyse reply and comment tweet text using natural language-based discourse analysis methods. 

During the $2019$ Alberta provincial and $2019$ Canadian federal elections, ParityBOT highlighted that hate speech is prevalent and difficult to combat on our social media platforms as they currently exist, and it is impacting democratic health and gender equality in our communities \cite{citron2009cyber}. 
We strategically designed ParityBOT to inject hope and positivity into politics, to encourage more diverse candidates to participate. By using machine learning technology to address these systemic issues, we can help change the discourse an link progress in science to progress in humanity.






\clearpage

\printbibliography

@article{rheault2019politicians,
  title={Politicians in the line of fire: Incivility and the treatment of women on social media},
  author={Rheault, Ludovic and Rayment, Erica and Musulan, Andreea},
  journal={Research \& Politics},
  volume={6},
  number={1},
  year={2019},
  publisher={SAGE Publications Sage UK: London, England}
}

@article{delisle2019large,
  title={A large-scale crowdsourced analysis of abuse against women journalists and politicians on Twitter},
  author={Delisle, Laure and Kalaitzis, Alfredo and Majewski, Krzysztof and de Berker, Archy and Marin, Milena and Cornebise, Julien},
  journal={arXiv preprint arXiv:1902.03093},
  year={2019}
}

@article{greenwood2019online,
  title={Online Abuse of UK MPs from 2015 to 2019},
  author={Greenwood, Mark A and Bakir, Mehmet E and Gorrell, Genevieve and Song, Xingyi and Roberts, Ian and Bontcheva, Kalina},
  journal={arXiv preprint arXiv:1904.11230},
  year={2019}
}

@inproceedings{golbeck2017large,
  title={A large labeled corpus for online harassment research},
  author={Golbeck, Jennifer and Ashktorab, Zahra and Banjo, Rashad O and Berlinger, Alexandra and Bhagwan, Siddharth and Buntain, Cody and Cheakalos, Paul and Geller, Alicia A and Gergory, Quint and Gnanasekaran, Rajesh Kumar and others},
  booktitle={Proceedings of the 2017 ACM on Web Science Conference},
  pages={229--233},
  year={2017},
  organization={ACM}
}

@inproceedings{hutto2014vader,
  title={Vader: A parsimonious rule-based model for sentiment analysis of social media text},
  author={Hutto, Clayton J and Gilbert, Eric},
  booktitle={8th AAAI on weblogs and social media},
  year={2014}
}

@book{trimble2018ms,
  title={Ms. Prime Minister: Gender, Media, and Leadership},
  author={Trimble, Linda},
  year={2018},
  publisher={U. Toronto Press}
}

@misc{toxictwitter,
    author = "Amnesty International",
    title = "Toxic Twitter: Women’s experiences of violence and abuse on Twitter",
    year = 2018,
    note = "\url{http://bit.ly/2jZTb5w}"
}

@misc{ipu,
    author = "Inter-Parliamentary Union",
    title = "Sexism, harassment and violence against women parliamentarians",
    year = 2018,
    month = 10,
    note = "\url{https://www.ipu.org/file/2425/download?token=0H5YdXVB}"
}

@article{phillips1998democracy,
  title={Democracy and representation: Or, why should it matter who our representatives are?},
  author={Phillips, Anne},
  journal={Feminism and politics},
  volume={224},
  pages={240},
  year={1998},
  publisher={Oxford University Press Oxford}
}

@inproceedings{he2008adasyn,
  title={ADASYN: Adaptive synthetic sampling approach for imbalanced learning},
  author={He, Haibo and Bai, Yang and Garcia, Edwardo A and Li, Shutao},
  booktitle={2008 IEEE International Joint Conference on Neural Networks (IEEE World Congress on Computational Intelligence)},
  pages={1322--1328},
  year={2008},
  organization={IEEE}
}

@incollection{olson2019tpot,
  title={TPOT: A tree-based pipeline optimization tool for automating machine learning},
  author={Olson, Randal S and Moore, Jason H},
  booktitle={Automated Machine Learning},
%   pages={151--160},
  year={2019},
  publisher={Springer}
}

@book{krug2009rocket,
  title={Rocket surgery made easy: The do-it-yourself guide to finding and fixing usability problems},
  author={Krug, Steve},
  year={2009},
  publisher={New Riders}
}

@article{citron2009cyber,
  title={Cyber civil rights},
  author={Citron, Danielle Keats},
  journal={BUL Rev.},
  volume={89},
  pages={61},
  year={2009},
  publisher={HeinOnline}
}

@inproceedings{ke2017lightgbm,
  title={Lightgbm: A highly efficient gradient boosting decision tree},
  author={Ke, Guolin and Meng, Qi and Finley, Thomas and Wang, Taifeng and Chen, Wei and Ma, Weidong and Ye, Qiwei and Liu, Tie-Yan},
  booktitle={Advances in Neural Information Processing Systems},
  pages={3146--3154},
  year={2017}
}

@misc{tweepy,
  title = {Tweepy},
  howpublished = {\url{http://www.tweepy.org/}},
  note = {Acc: 2019-09-06}
}

@misc{twitterdev,
  title = {Twitter Developer Agreement and Policy},
  howpublished = {\url{https://developer.twitter.com/en/developer-terms/agreement-and-policy.html}},
  note = {Acc: 2019-09-06}
}

@misc{perspective,
  title = {Perspective API},
  howpublished = {\url{https://www.perspectiveapi.com/}},
  note = {Acc: 2019-09-06}
}

@misc{hatesonar,
  title = {HateSonar},
  howpublished = {\url{https://github.com/Hironsan/HateSonar}},
  note = {Acc: 2019-09-06}
}

@misc{wiki,
  title = {Wikipedia: List of candidates by riding for the 43rd Canadian federal election},
  howpublished = {\url{https://w.wiki/7zT }},
  note = {Acc: 2019-09-06}
}

@misc{gender,
  title = {Gender Guesser},
  howpublished = {\url{https://pypi.org/project/gender-guesser/}},
  note = {Acc: 2019-09-06}
}

@misc{toxiccard,
    title = {Toxicity Model Card},
    howpublished = {\url{https://github.com/conversationai/perspectiveapi/blob/master/2-api/model-cards/English/toxicity.md}},
    note = {Acc: 2019-11-14}
}

@inproceedings{mitchell2019model,
  title={Model cards for model reporting},
  author={Mitchell, Margaret and Wu, Simone and Zaldivar, Andrew and Barnes, Parker and Vasserman, Lucy and Hutchinson, Ben and Spitzer, Elena and Raji, Inioluwa Deborah and Gebru, Timnit},
  booktitle={Proceedings of the Conference on Fairness, Accountability, and Transparency},
  pages={220--229},
  year={2019},
  organization={ACM}
}

@article{Africaworkshop,
  title={OHCHR expert workshops on the prohibition of incitement to national, racial or religious hatred},
  author={Heiner Bielefeldt, Frank La Rue, and Githu Muigai},
  journal={Expert workshop on Africa},
  year={2011}
}

@inproceedings{waseem2016hateful,
  title={Hateful symbols or hateful people? predictive features for hate speech detection on twitter},
  author={Waseem, Zeerak and Hovy, Dirk},
  booktitle={Proceedings of the NAACL student research workshop},
  pages={88--93},
  year={2016}
}

@inproceedings{basile2019semeval,
  title={Semeval-2019 task 5: Multilingual detection of hate speech against immigrants and women in twitter},
  author={Basile, Valerio and Bosco, Cristina and Fersini, Elisabetta and Nozza, Debora and Patti, Viviana and Pardo, Francisco Manuel Rangel and Rosso, Paolo and Sanguinetti, Manuela},
  booktitle={Proceedings of the 13th International Workshop on Semantic Evaluation},
  pages={54--63},
  year={2019}
}

@inproceedings{founta2018large,
  title={Large scale crowdsourcing and characterization of twitter abusive behavior},
  author={Founta, Antigoni Maria and Djouvas, Constantinos and Chatzakou, Despoina and Leontiadis, Ilias and Blackburn, Jeremy and Stringhini, Gianluca and Vakali, Athena and Sirivianos, Michael and Kourtellis, Nicolas},
  booktitle={Twelfth International AAAI Conference on Web and Social Media},
  year={2018}
}

\clearpage
\setcounter{section}{0}

\section{Tweet Cleaning and Feature Details}

\subsection{Tweet Cleaning Methods}
\label{appdx:clean}

We use regular expression rules to clean tweets: convert the text to lowercase, remove URLs, strip newlines, replace whitespace with a single space, and replace mentions with the text tag `MENTION'. While these rules may bias the classifiers, they allow for consistency and generalization between training, validation, and testing datasets.

\subsection{Tweet Featurization Details}
\label{appdx:featz}


Each tweet is processed by three models: Perspective API from Jigsaw \cite{perspective}, HateSonar \cite{hatesonar}, and VADER sentiment models \cite{hutto2014vader}. Each of these models outputs a score between $[0,1]$ which correlates the text of the tweet with the specific measure of the feature. The outputs from these models (i.e. $17$ from Perspective, $3$ from HateSonar, and $4$ from VADER) are combined into a single feature vector for each tweet. Below we list the outputs for each text featurization model:

\setlist{nolistsep}
\begin{itemize}[noitemsep]
    \item \textbf{Perspective API}: 'IDENTITY\_ATTACK', 'INCOHERENT', 'TOXICITY\_FAST', 'THREAT', 'INSULT', 'LIKELY\_TO\_REJECT', 'TOXICITY', 'PROFANITY', 'SEXUALLY\_EXPLICIT', 'ATTACK\_ON\_AUTHOR', 'SPAM', 'ATTACK\_ON\_COMMENTER', 'OBSCENE', 'SEVERE\_TOXICITY', 'INFLAMMATORY'
    \item \textbf{HateSonar}: 'sonar\_hate\_speech', 'sonar\_offensive\_language', 'sonar\_neither'
    \item \textbf{VADER}: 'vader\_neg', 'vader\_neu', 'vader\_pos', 'vader\_compound'
\end{itemize}


\subsection{Validation and Ablation Experiments}
\label{appdx:ablate}

For validation, we found the most relevant features and set an abusive prediction threshold by using a dataset of $20194$ cleaned, unique tweets identified as either hateful and not hateful from previous research \cite{golbeck2017large}. Each entry in our featurized dataset is composed of $24$ features and a class label of \textit{hateful or not hateful}. The dataset is shuffled and randomly split into training (80\%) and testing (20\%) sets matching the class balance ($25.4\%$ hateful) of the full dataset. We use Adaptive Synthetic (ADASYN) sampling to resample and balance class proportions in the dataset \cite{he2008adasyn}. 

With the balanced training dataset, we found the best performing classifier to be a gradient boosted decision tree \cite{ke2017lightgbm} by sweeping over a set of possible models and hyperparameters using TPOT \cite{olson2019tpot}. For this sweep, we used $10$-fold cross validation on the training data. We randomly partition this training data $10$ times, fit a model on a training fraction, and validate on the held-out set.

We performed an ablation experiment to test the relative impact of the features derived from the various text classification models.

\begin{figure}
    \centering
    \includegraphics[scale=0.4]{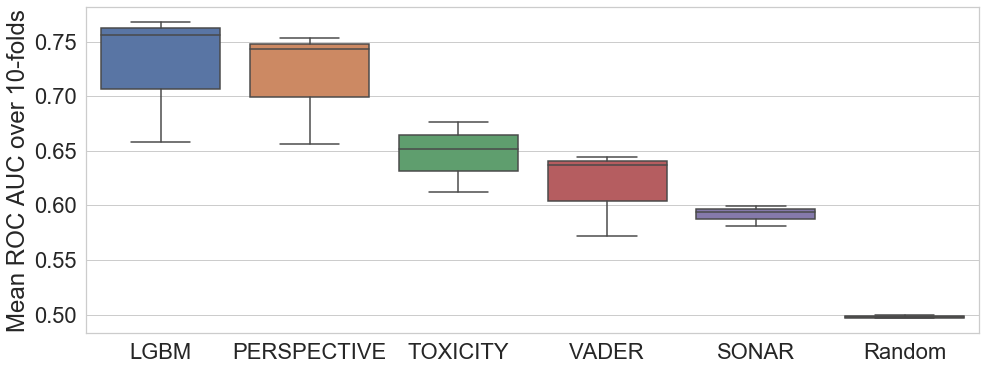}
    \caption{10-fold cross validation ablation experiment showing the relative impacts of including various feature sets (\textsc{Perspective} \cite{perspective}, \textsc{Vader} \cite{hutto2014vader}, \textsc{HateSonar} \cite{hatesonar}, from left to right) in the  feature vectors on performance. Performance is measured using binary classification area under the receiver operated characteristic curve (ROC AUC) and averaged over the 10-folds. These methods are compared with the best performing classifier from the validation study, a gradient-boosted decision tree (textsc{LGBM} \cite{ke2017lightgbm}, left) and a stratified \textsc{RANDOM} classifier (right).}
    \label{fig:my_label}
\end{figure}

\section{Quantitative analysis of elections}
\label{appdx:results}

This table includes quantitative results from the deployment of ParityBOT in the Alberta 2019 provincial and Canadian 2019 federal elections.

\begin{tabular}{ | c | c | c | }
 \hline
\makecell{} & \makecell{\textbf{Alberta 2019 provincial election} \\ Apr. 1-15 2019} & \makecell{\textbf{Canada 2019 federal election} \\ Sep. 11-Oct. 26 2019} \\ \hline
Total positivitweets sent & 973 & 2428 \\ \hline
Total impressions & 84,961 & 304,600 \\ \hline
Total retweets & 142 & 529 \\ \hline
Total likes & 412 & 1500 \\ \hline
Total replies & n/a & 30 \\ \hline
Total tweets analysed & 12,726 & 228,255 \\ \hline
Total tweets scored abusive & 1468 & 9987 \\ \hline
Abusive rate & 7.65\% & 4.38\% \\ \hline
Total candidates tracked & 90 & 314 \\ \hline
Decision threshold & 0.8 (80\% likely to be abusive) & 0.9 (90\% likely to be abusive) \\ \hline
\end{tabular}

\section{ParityBOT Research Plan and Discussion Guide}

\textit{Overview} Interviews will be completed in three rounds with three different target participant segments.

\textit{Research Objectives}
\setlist{nolistsep}
\begin{itemize}[noitemsep]
    \item Understand if and how the ParityBOT has impacted women in politics
    \item Obtain feedback from Twitter users who've interacted with the bot 
    \item Explore potential opportunities to build on the existing idea and platform
    \item Gain feedback and initial impressions from people who haven't interacted with the Bot, but are potential audience 
\end{itemize} 

\textit{Target Participants}

\setlist{nolistsep}
\begin{itemize}[noitemsep]
    \item Round 1: Women in politics who are familiar with the Bot 
    \item Round 2: Women who've interacted with the Bot (maybe those we don't know) 
    \item Round 3: Some women who may be running in the federal election who haven't heard of the ParityBOT, but might benefit from following it
    \item All participants: 
    Must be involved in politics in Canada and must be engaged on Twitter - i.e. have an account and follow political accounts and/or issues
\end{itemize}

\textit{Recruiting}

\setlist{nolistsep}
\begin{itemize}[noitemsep]
    \item Round 1: [Author] recruit from personal network via text
    \item Round 2: Find people who've interacted with the bot on Twitter who we don't know, send them a DM, and ask if we can get their feedback over a 15- to 30-minute phone call 
    \item Round 3: Use contacts in Canadian politics to recruit participants who have no prior awareness of ParityBOT
\end{itemize}

\textit{Method} 15- to 30-minute interviews via telephone 

\textit{Output} Summary of findings in the form of a word document that can be put into the paper

\subsection{Discussion Guide}
\label{appdx:disc}

\textbf{Introduction}

\textit{[Author]:}
Hey! Thanks for doing this. This shouldn't take longer than 20 minutes. [Author] is a UX researcher and is working with us. They'll take it from here and explain our process, get your consent and conduct the interview. I'll be taking notes. Over to [Author]! 

\textit{[Author]:}
Hi, my name is [Author], I'm working with [Author] and [Author] to get feedback on the ParityBOT; the Twitter Bot they created during the last provincial election. 

With your permission, we'd like to record our conversation. The recording will only be used to help us capture notes from the session and figure out how to improve the project, and it won't be seen by anyone except the people working on this project. We may use some quotes in an academic paper, You'll be anonymous and we won't identify you personally by name. 

If you have any concerns at time, we can stop the interview and the recording. Do we have your permission to do this? (Wait for verbal ``\textbf{yes}'').

\textbf{Round 1 (Women in Politics familiar with ParityBOT)}

\textit{Background and Warm Up}

\setlist{nolistsep}
\begin{itemize}[noitemsep]
    \item When you were thinking about running for politics what were your major considerations? For example, barriers, concerns?
    \item We know that online harassment is an issue for women in politics - have you experienced this in your career? How do you deal with harassment? What are your coping strategies? 
    \item What advice would you give to women in politics experiencing online harassment? 
\end{itemize}

\textit{Introduction to PartyBOT} Thanks very much, now, more specifically about the ParityBOT:

\setlist{nolistsep}
\begin{itemize}[noitemsep]
    \item What do you know about the ParityBOT? 
    \item What do you think it's purpose is? 
    \item Did you encounter it? Tell me about how you first encountered it? Did it provide any value to you during your campaign? How? Do you think this is a useful tool? Why or why not? Did it mitigate the barrier of online harassment during your time as a politician? 
    \item Is there anything you don't like about the Bot? 
\end{itemize}

\textit{Next Steps} If you could build on this idea of mitigating online harassment for women in politics, what ideas or suggestions would you have? 

\textit{Conclusion} Any other thoughts or opinions about the ParityBOT you'd like to share before we end our call? 

Thank you very much for your time! If you have any questions, or further comments, feel free to text or email [Author].

\end{document}